\documentstyle[12pt]{article}
\vspace{1.8cm} \oddsidemargin 0pt \evensidemargin 0pt  \topmargin
-1.5cm \footheight 26pt \footskip 10mm

\begin{document}
\baselineskip 20pt
\begin{center}
\baselineskip=24pt {\Large\bf Classical and quantum theory for
Superluminal particle}

\vspace{1cm} \centerline{Xiang-Yao Wu$^{a}$
\footnote{E-mail:wuxy2066@163.com}}Xiao-Jing Liu$^{a}$, Li
Wang$^{a}$, Yi-Qing Guo $^{b}$ and Xi-Hui Fan$^{c}$ \vspace{0.8cm}

\noindent{\footnotesize a. \textit{Institute of Physics,
Jilin Normal University, Siping 136000, China}}\\
{\footnotesize b. \textit{Institute of High Energy Physics,
P.O.Box 918(4), Beijing 100039, China}}\\
{\footnotesize c. \textit{Department of Physics, Qufu Normal
University, Qufu 273165, China}}

\end{center}
\date{}
\renewcommand{\thesection}{Sec. \Roman{section}} \topmargin 10pt
\renewcommand{\thesubsection}{ \arabic{subsection}} \topmargin 10pt
{\vskip 5mm
\begin {minipage}{140mm}
\centerline {\bf Abstract} \vskip 8pt
\par

\indent\\

\hspace{0.3in}As we all know, when the relative velocity of two
inertial reference frames $\sum$ and $\sum^{'}$ is less than the
speed of light, the relations of $x_{\mu}$ with $x_{\mu}^{'}$, a
particle mass $m$ with its velocity $\upsilon$, and a particle
mass with its energy are all given by Einstein's special
relativity. In this paper, we will give new relation of $x_{\mu}$
and $x_{\mu}^{'}$ when the relative velocity of $\sum$ and
$\sum^{'}$ frame is larger than the speed of light, and also we
give the relation of a particle mass $m$ with its velocity
$\upsilon$, and a particle mass $m$ with its energy $E$ when the
particle velocity $v$ is larger than the speed of light.

\end {minipage}

\newpage

{\bf 1. Introduction} \vskip 8pt A hundred years ago, Einstein
laid the foundation for a revolution in our conception of space
and time, matter and energy. Later, special theory of relativity
was accepted by mainstream physicists. It is  based on two
postulates by Einstein \cite{s1}:

1. The Principle of Relativity: All laws of nature are the same in
all inertial reference frames. In other words, we can say that the
equation expressing the laws of nature are invariant with respect
to transformations of coordinates and time from one inertial
reference frame to another.

2. The Universal Speed of Light: The speed of light in vacuum is
the same for all inertia observers, regardless of the motion of
the source, the observer, or any assumed medium of propagation.

The invariant principle of the speed of light is right in all
inertial reference frames in which their relative velocity
$\upsilon$ is less than the speed of light $c$. Since the light
velocity has no relation with the movement of light source, which
has been proved by experiment \cite{s2}, we can consider a moving
light source as a inertial reference frames, and we can obtain the
result that the speed of light has nothing to do with the movement
speed of the inertial reference frames, i.e., the speed of light
is same in all inertial reference frames. It derives that light
has no interaction with light source, and the light has no
inertia. So, the rest mass of light tend to zero. Recently, a
series of experiments have revealed that electromagnetic wave was
able to travel at a group velocity faster than $c$. These
phenomena have been observed in dispersive media [3-4], in
electronic circuits \cite{s5}, and in evanescent wave cases
\cite{s6}. In fact, over the last decade, the discussion of the
tunnelling time problem has experienced a new stimulus by the
results of analogous experiments with evanescent electromagnetic
wave packets \cite{s7}, and the superluminal effects of evanescent
waves have been revealed in photon tunnelling experiments in both
the optical domain and the microwave range \cite{s6}. In nature,
maybe there is superluminal phenomena, and the relative velocity
$\upsilon$ of two inertial reference frames can be larger than $c$
or equal to $c$. The superluminal phenomena can also appear in the
progress of light propagation. For example, when a beam of light
move at the same direction, all photons relative velocity
$\upsilon$ is equal to zero and not the speed of light $c$. So,
the postulates about the invariant principle of the speed of light
is incorrect when the relative velocity of the two inertial
reference frames is equal to the speed of light (we regard the
light as a inertial reference frames). The result that the speed
of light $c$ is maximum speed is also incorrect, because the
relative velocity of two beams of light which move along opposite
direction exceeds the speed of light $c$. So, when two inertial
reference frames relative velocity $\upsilon$ is larger than the
speed of light or equal to the speed of light, Einstein's
postulation of the invariant principle of the speed of light
should be modified. We think that there are two ranges of velocity
in nature: One is in the range of $0\leq \upsilon < c$, which is
suit for special relativity. The other is in the range of $c\leq
\upsilon < c_{m}$ ($c_{m}$ is the maximum velocity in nature),
which will be researched in this paper. \\

{\bf 2. The space-time transformation and mass-energy relation of
special relativity} \vskip 8pt In 1905, Einstein gave the
space-time transformation and mass-energy relation which are based
on his two postulates. The space-time transformation is
\begin{eqnarray}
x&=&\frac{x^{'}+\upsilon
t^{'}}{\sqrt{1-\frac{\upsilon^{2}}{c^{2}}}}\nonumber\\
 y&=&y^{'}\nonumber\\
z&=&z^{'}\nonumber\\
t&=&\frac{t^{'}+\frac{\upsilon}{c^{2}}x^{'}}{\sqrt{1-\frac{\upsilon^{2}}{c^{2}}}},
\end{eqnarray}
where $x$, $y$, $z$, $t$ are space-time coordinates in $\sum$
frame, $x^{'}$, $y^{'}$, $z^{'}$, $t^{'}$ are space-time
coordinates in $\sum^{'}$ frame, $\upsilon$ is the relative
velocity that $\sum$ and $\sum^{'}$ frame move along with $x$ and
$x^{'}$ axes, and $c$ is the speed of light.
The velocity transformation is\\
\begin{eqnarray}
u_{x}&=&\frac{u_{x}^{'}+\upsilon
}{1+\frac{\upsilon u_{x}^{'}}{c^{2}}}\nonumber\\
u_{y}&=&\frac{u_{y}^{'}\sqrt{1-\frac{\upsilon^{2}}{c^{2}}}}
{1+\frac{\upsilon u_{x}^{'}}{c^{2}}}\nonumber\\
u_{z}&=&\frac{u_{z}^{'}\sqrt{1-\frac{\upsilon^{2}}{c^{2}}}}
{1+\frac{\upsilon u_{x}^{'}}{c^{2}}},
\end{eqnarray}
where $u_{x}$, $u_{y}$ and $u_{z}$ are a particle velocity in
$\sum$ frames, $u_{x}^{'}$, $u_{y}^{'}$ and $u_{z}^{'}$ are the
particle velocity in $\sum^{'}$ frames. The relation of a particle
mass $m$ with its movement velocity $\upsilon$ is
\begin{equation}
m=\frac{m_{0}}{\sqrt{1-\frac{\upsilon^{2}}{c^{2}}}},
\end{equation}
with $m_{0}$ and $m$ being the particle rest mass and relativistic
mass. The relation of a particle relativistic energy $E$ with its
relativistic mass $m$ is
\begin{equation}
E=mc^{2},
\end{equation}
and the relation of particle energy $E$ with its momentum $p$ is
\begin{equation}
E^{2}=m_{0}^{2}c^{4}+p^{2}c^{2}.
\end{equation}

{\bf 3. The space-time transformation for superluminal reference
frames}

 \vskip 8pt

At some 40 years ago, O.M.P. Bilaniuk, V.K. Deshpande and E.S.G.
Sudarshan had researched the space-time relation for superluminal
reference frames within the framework of special relativity [8,
9]. They thought the space-time and velocity transformation of
special relativity are suit for superluminal reference frames, and
they obtained the new results that the proper length $L_{0}$ and
proper time $T_{0}$ must be imaginary so that the measured
quantities length $L$ and time $T$ are real. In the following, we
will give the relation of space-time in two inertial reference
frames $\sum$ and $\sum^{'}$ when their relative velocity
$\upsilon$ is larger than the speed of light $c$. We think even if
there is superluminal movement, the movement velocity can not be
infinity. So, we can think there is a limit velocity in nature,
which is called maximum velocity $c_{m}$. All particles movement
velocity can not exceed the maximum velocity $c_{m}$ in arbitrary
inertial reference frames. In the velocity range of $c\leq
\upsilon <c_{m}$, we give two postulates as follows:

1. The Principle of Relativity: All laws of nature are the same in
all inertial reference frames.

2. The Universal of Maximum Velocity: There is a maximum velocity
$c_{m}$ in nature, and the $c_{m}$ is invariant in all inertial
reference frames.

From the two postulates, we can obtain the space-time
transformation and velocity transformation, which are similar as
Lorentz transformation of special relativity:
\begin{eqnarray}
x&=&\frac{x^{'}+\upsilon
t^{'}}{\sqrt{1-\frac{\upsilon^{2}}{c^{2}_{m}}}}\nonumber\\
 y&=&y^{'}\nonumber\\
z&=&z^{'}\nonumber\\
t&=&\frac{t^{'}+\frac{\upsilon}{c^{2}_{m}}x^{'}}{\sqrt{1-\frac{\upsilon^{2}}{c^{2}_{m}}}},
\end{eqnarray}
and
\begin{eqnarray}
u_{x}&=&\frac{u_{x}^{'}+\upsilon
}{1+\frac{\upsilon u_{x}^{'}}{c^{2}_{m}}}\nonumber\\
u_{y}&=&\frac{u_{y}^{'}\sqrt{1-\frac{\upsilon^{2}}{c^{2}_{m}}}}
{1+\frac{\upsilon u_{x}^{'}}{c^{2}_{m}}}\nonumber\\
u_{z}&=&\frac{u_{z}^{'}\sqrt{1-\frac{\upsilon^{2}}{c^{2}_{m}}}}
{1+\frac{\upsilon u_{x}^{'}}{c^{2}_{m}}},
\end{eqnarray}
where $\upsilon$ $(v\geq c)$ is the relative velocity of $\sum$
and $\sum^{'}$ frame. Now, We can discuss the problem of the speed
of light from Eq. (7). For two inertial reference frames $\sum$
and $\sum^{'}$, the $\sum^{'}$ frame is a rest frame for light,
i.e., their relative velocity $v$ is equal to $c$. At the time
$t=0$, a beam of light are emitted from origin $O$. From Eq. (7),
we have
\begin{equation}
u_{x}=c,
\end{equation}
then
\begin{equation}
u_{x}^{'}=0,
\end{equation}
and
\begin{equation}
u_{x}=-c,
\end{equation}
and then
\begin{equation}
u_{x}^{'}=\frac{-c-c}{1+\frac{c^{2}}{c_{m}^{2}}}=-2\frac{c
c_{m}^{2}}{c^{2}+c_{m}^{2}}>-2c.
\end{equation}
It show that the invariant principle of the speed of light is
violated in the frame of the speed of light.

{\bf 4. The relation of mass with velocity for superluminal
particle}

In Refs. [8, 9], the authors thought that the relation of
particle's mass with its velocity and energy with its mass in
special relativity are also suit for superluminal particles, and
they obtained the interesting result that the rest mass of
particle $m_{0}$ must be imaginary so that the the particle's
energy and momentum are real. In the following, we will give the
new relation of particle mass $m$ with its velocity $v$. We can
consider the collision between two identical particle. It is shown
in Fig. 1

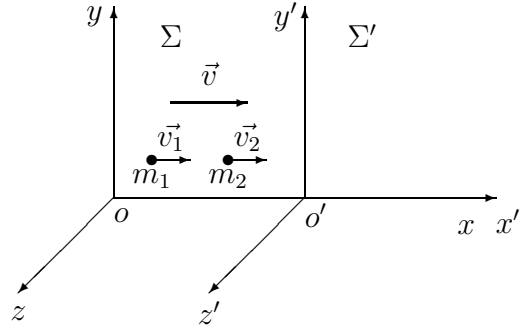
\setlength{\unitlength}{0.1in}

\begin{picture}(100,15)
\put(20,4){\vector(1,0){20}}
  \put(38,2){\makebox(2,1)[l]{$x$}}
  \put(40,2.3){\makebox(2,1)[l]{$x^{\prime}$}}
\put(23,9){\vector(1,0){4}}
  \put(24,10){\makebox(2,1)[c]{$\vec{v}$}}
\put(22,6){\vector(1,0){2}}
  \put(22,6.7){\makebox(2,1)[c]{$\vec{v_{1}}$}}
  \put(21,4.5){\makebox(2,1)[c]{$m_{1}$}}
\put(26,6){\vector(1,0){2}}
  \put(26,6.7){\makebox(2,1)[c]{$\vec{v_{2}}$}}
  \put(25,4.5){\makebox(2,1)[c]{$m_{2}$}}
\put(20,4){\vector(0,1){10}}
  \put(18,13){\makebox(2,1)[c]{$y$}}
\put(30,4){\vector(0,1){10}}
  \put(28,13){\makebox(2,1)[c]{$y^{\prime}$}}
\put(30,4){\vector(-1,-1){5}}
  \put(24,-2.5){\makebox(2,1)[c]{$z^{\prime}$}}
\put(20,4){\vector(-1,-1){5}}
  \put(14,-2.5){\makebox(2,1)[c]{$z$}}

  \put(30,2.6){\makebox(2,1)[l]{$o^{\prime}$}}
  \put(20,2.6){\makebox(2,1)[l]{$o$}}
  \put(32,12){\makebox(2,1)[c]{$\Sigma^{\prime}$}}
  \put(22,12){\makebox(2,1)[c]{$\Sigma$}}
 \put(22,6){\circle*{0.5}}
 \put(26,6){\circle*{0.5}}
\end{picture}

\vskip 15pt

The $\Sigma$ is laboratory system, and $\Sigma^{\prime}$ is
mass-center system of tow particles. In $\Sigma$ system, the
velocity of tow particles $m_{1}$ and $m_{2}$ are $\vec{v_{1}}$
and $\vec{v_{2}}$ $(v_{1}> v_{2}\geq c)$, which are along with $x
(x^{\prime})$ axis, and they are $v^{\prime}$ and $-v^{\prime}$ in
$\Sigma^{\prime}$ system. After collision, the velocity of tow
particles are all $v$ $(v\geq c)$ in $\Sigma$ system. Momentum was
conserved in this process:
\begin{equation}
m_{1}v_{1}+m_{2}v_{2}=(m_{1}+m_{2})v.
\end{equation}
According to equation (7),
\begin{eqnarray}
v_{1}=\frac{v^{\prime}+v}{1+\frac{v v^{\prime}}{c_{m}^{2}}} \nonumber\\
v_{2}=\frac{-v^{\prime}+v}{1-\frac{v v^{\prime}}{c_{m}^{2}}}.
\end{eqnarray}
From equations (12) and (13), we get
\begin{equation}
m_{1}(1-\frac{v v^{\prime}}{c_{m}^{2}})=m_{2}(1+\frac{v
v^{\prime}}{c_{m}^{2}}),
\end{equation}
from equation (7), we can obtain
\begin{equation}
1+\frac{u_{x}^{\prime}v}{c_{m}^{2}}=\frac{\sqrt{1-\frac{{u^{\prime}}^{
2}}{c_{m}^{2}}} \sqrt{1-\frac{v^{2}}{c_{m}^{2}}}}
{\sqrt{1-\frac{u^{2}}{c_{m}^{2}}}},
\end{equation}
for the particle $m_{1}$, the equation (15) is
\begin{equation}
1+\frac{v^{\prime}v}{c_{m}^{2}}=\frac{\sqrt{1-\frac{v^{\prime2}}{c_{m}^{2}}}
\sqrt{1-\frac{v^{2}}{c_{m}^{2}}}}
{\sqrt{1-\frac{v_{1}^{2}}{c_{m}^{2}}}},
\end{equation}
for particle $m_{2}$, the equation (15) is
\begin{equation}
1-\frac{v^{\prime}v}{c_{m}^{2}}=\frac{\sqrt{1-\frac{v^{\prime2}}{c_{m}^{2}}}
\sqrt{1-\frac{v^{2}}{c_{m}^{2}}}}
{\sqrt{1-\frac{v_{2}^{2}}{c_{m}^{2}}}}.
\end{equation}
On substituting equations (16) and (17) into (14), we get
\begin{equation}
m(v_{1})\sqrt{1-\frac{v_{1}^{2}}{c_{m}^{2}}}=
m(v_{2})\sqrt{1-\frac{v_{2}^{2}}{c_{m}^{2}}}=
m(c)\sqrt{1-\frac{c^{2}}{c_{m}^{2}}}=constant,
\end{equation}
where $m(c)$ is the particle mass when when its velocity is equal
to $c$.

For velocity $v$ ($v>c$), we have
\begin{equation}
m(v)\sqrt{1-\frac{v^{2}}{c_{m}^{2}}}=m(c)\sqrt{1-\frac{c^{2}}{c_{m}^{2}}},
\end{equation}
and hence
\begin{equation}
m(v)=m_{c}\sqrt{\frac{c_{m}^{2}-c^{2}}{c_{m}^{2}-v^{2}}}.
\end{equation}
with $m_{c}=m(c)$. The equation (20) is the relation of
superluminal particle mass $m$ with its velocity $v$ ($v\geq c$).

{\bf 5. The relation of energy with mass for superluminal
particle}

In the following, we define a 4-vector of space-time
\begin{equation}
x_{\mu}=(x_{1}, x_{2}, x_{3}, x_{4})=(x, y, z, ic_{m}t).
\end{equation}
The invariant interval $ds^{2}$ is given by the equation
\begin{equation}
ds^{2}=-dx_{\mu}dx_{\mu}=c_{m}^{2}dt^{2}-(dx)^{2}-(dy)^{2}-(dz)^{2}=c_{m}^{2}d\tau^{2},
\end{equation}
we get
\begin{equation}
d\tau=\frac{1}{c_{m}}ds,
\end{equation}
where $d\tau$ is proper time, the 4-velocity can be defined by
\begin{equation}
U_{\mu}=\frac{dx_{\mu}}{d\tau}=\frac{dx_{\mu}}{dt}\frac{dt}{d\tau}=\gamma_{\mu}(\vec{v},
ic_{m}),
\end{equation}
where $\gamma_{\mu}=\frac{1}{\sqrt{1-\frac{v^{2}}{c_{m}^{2}}}}$,
$\vec{v}=\frac{d\vec{x}}{dt}$ and $\frac{dt}{d\tau}=\gamma_{\mu}$.
We can define 4-momentum as
\begin{equation}
p_{\mu}=m_{c}U_{\mu}=(\vec{p}, ip_{4}),
\end{equation}
with
$\vec{p}=\frac{m_{c}\vec{v}}{\sqrt{1-\frac{v^{2}}{c_{m}^{2}}}}$,
$p_{4}=\frac{m_{c}c_{m}}{\sqrt{1-\frac{v^{2}}{c_{m}^{2}}}}$. We
can define a particle energy as
\begin{equation}
E=\frac{m_{c}c_{m}^{2}}{\sqrt{1-\frac{v^{2}}{c_{m}^{2}}}},
\end{equation}
then
\begin{equation}
p_{\mu}=(\vec{p}, \frac{i}{c_{m}}E),
\end{equation}
the invariant quantity constructed from this 4-vector is
\begin{equation}
p_{\mu}p_{\mu}=p^{2}-\frac{E^{2}}{c_{m}^{2}}=-m_{c}^{2}c_{m}^{2},
\end{equation}
i.e.,
\begin{equation}
E^{2}-p^{2}c_{m}^{2}=m_{c}^{2}c_{m}^{4}.
\end{equation}
The equation (29) is the relation of superluminal particle mass
$m$, momentum $\vec{p}$ and energy $E$.

On substituting (20) into (26), we can obtain the relation of
mass-energy for superluminal particle.
\begin{equation}
E=\frac{m(v)}{\sqrt{c_{m}^{2}-c^{2}}}c_{m}^{3}.
\end{equation}

In the following, we research the problem of photon mass. From Eq.
(26), we can obtain the mass of photon when the photon velocity is
$c$.
\begin{equation}
E_{\nu}=\frac{m_{c\nu}c_{m}^{2}}
{\sqrt{1-\frac{c^{2}}{c_{m}^{2}}}}=h\nu,
\end{equation}

i.e.,
\begin{equation}
m_{c\nu}=\frac{h\nu\sqrt{1-\frac{c^{2}} {c_{m}^{2}}}}{c_{m}^{2}}.
\end{equation}
If there is a superluminal photon, we can calculate its mass and
energy. On substituting equation (32) into (20), we can obtain the
mass of the superluminal photon
\begin{equation}
m_{\nu}=m_{c \nu}\sqrt{\frac{c_{m}^{2}-c^{2}}{c_{m}^{2}-v^{2}}}
=\frac{h\nu}{c_{m}^{3}}\frac{c_{m}^{2}-c^{2}}
{\sqrt{c_{m}^{2}-v^{2}}}  \hspace{0.3in}  (v>c),
\end{equation}
from equations (30) and (33), we can obtain the energy of the
superluminal photon
\begin{equation}
E=\frac{m_{\nu}c_{m}^{3}}{\sqrt{c_{m}^{2}-c^{2}}}=h\nu
\sqrt{\frac{c_{m}^{2}-c^{2}}{c_{m}^{2}-v^{2}}}=h\nu^{\prime},
\end{equation}
where $\nu^{\prime}$ is the frequency of superluminal photon. It
is
\begin{equation}
\nu^{\prime}=\nu\sqrt{\frac{c_{m}^{2}-c^{2}}{c_{m}^{2}-v^{2}}}>
\nu.
\end{equation}
It is shown that the frequency of superluminal photon is larger
than the frequency of light velocity photon.

{\bf 6. The relativistic dynamics for superluminal particle}

In the following, we research the relativistic dynamics for
superluminal particle. We define a 4-force as:
\begin{equation}
K_{\mu}=\frac{dp_{\mu}}{d\tau},
\end{equation}
from equation (27), we have
\begin{equation}
K_{\mu}=(\vec{K}, iK_{4}),
\end{equation}
the "ordinary" force $\vec{K}$ is
\begin{equation}
\vec{K}=\frac{d\vec{p}}{dt}
\frac{dt}{d\tau}=\frac{1}{\sqrt{1-\frac{v^{2}}{c_{m}^{2}}}}\frac{d\vec{p}}{dt},
\end{equation}
while the fourth component
\begin{eqnarray}
K_{4}&=&\frac{dp_{4}}{d\tau}=\frac{1}{c_{m}}\frac{dE}{d\tau} \nonumber\\
&=&\frac{1}{c_{m}}\frac{d}{d\tau}\sqrt{m_{c}^{2}c_{m}^{4}+p^{2}c_{m}^{2}}\nonumber\\
&=&c_{m}\frac{1}{E}\vec{p}\cdot \frac{d\vec{p}}{d\tau}\nonumber\\
&=&\frac{1}{c_{m}}\vec{v}\cdot \vec{K},
\end{eqnarray}
and so
\begin{equation}
K_{\mu}=(\vec{K}, \frac{i}{c_{m}}\vec{v}\cdot \vec{K}),
\end{equation}
the covariant equation for the superluminal particle are
\begin{equation}
\vec{K}=\frac{d\vec{p}}{d\tau},
\end{equation}
\begin{equation}
\vec{K}\cdot \vec{v}=\frac{dE}{d\tau},
\end{equation}
the equations (36) and (37) can be written as:
\begin{equation}
\vec{K}=\frac{d\vec{p}}{d\tau}=\frac{d\vec{p}}{dt}\frac{dt}{d\tau}
=\frac{d\vec{p}}{dt}\frac{1}{\sqrt{1-\frac{v^{2}}{c_{m}^{2}}}},
\end{equation}

\begin{equation}
\sqrt{1-\frac{v^{2}}{c_{m}^{2}}}\vec{K}=\frac{d\vec{p}}{dt},
\end{equation}

\begin{equation}
\sqrt{1-\frac{v^{2}}{c_{m}^{2}}}\vec{K}\cdot
\vec{v}=\frac{dE}{dt},
\end{equation}
we define force $\vec{F}$ as
\begin{equation}
\vec{F}=\sqrt{1-\frac{v^{2}}{c_{m}^{2}}}\vec{K}.
\end{equation}
The relativistic dynamics equation for the superluminal particle
are
\begin{equation}
\vec{F}=\frac{d\vec{p}}{dt},
\end{equation}
\begin{equation}
\vec{K}\cdot \vec{v}=\frac{dE}{dt}.
\end{equation}

{\bf 7. The quantum wave equation for superluminal particle}
\vskip 8pt In the following, we will give the quantization wave
equation for a superluminal particle.

Form equation (29), we express $E$ and $\vec{p}$ as operators:
\begin{eqnarray}
E\rightarrow i\hbar\frac{\partial}{\partial t} \nonumber\\
\vec{p}\rightarrow -i\hbar \nabla,
\end{eqnarray}
then we obtain the quantum wave equation of superluminal particle
\begin{equation}
[\frac{\partial^{2}}{\partial
t^{2}}-c_{m}^{2}\nabla^{2}+\frac{m_{c}^{2}c_{m}^{4}}{\hbar^{2}}]\Psi(\vec{r},
t)=0.
\end{equation}
The equation is similar as the K-G equation. At $m_{c}=0$, we have
\begin{equation}
[\frac{\partial^{2}}{\partial
t^{2}}-c_{m}^{2}\nabla^{2}]\Psi(\vec{r}, t)=0.
\end{equation}
The equation is similar as the wave equation of photon.

We know that the particle wave equation of spin $\frac{1}{2}$ is
Dirac equation
\begin{equation}
i\hbar \frac{\partial}{\partial t}\Psi=[-i\hbar c
\vec{\alpha}\cdot \vec{\nabla}+mc^{2}\beta]\Psi,
\end{equation}
where $\alpha$ and $\beta$ are matrixes
\[
\alpha= \left (
\begin {array} {cc}
0 & \vec{\sigma} \\
\vec{\sigma} & 0
\end{array} \right ),
\]
and
\[
\beta= \left (
\begin {array} {cc}
I & 0 \\
0 & -I
\end{array} \right ),
\]
where $\vec{\sigma}$ are Pauli matrixes, and $I$ is unit matrix of
$2\times 2$.

We can give the superluminal particle wave equation for spin
$\frac{1}{2}$, which it is similar as Dirac equation
\begin{equation}
i \hbar \frac{\partial}{\partial t}\Psi=[-i\hbar
c_{m}\vec{\alpha}\cdot \vec{\nabla}+m_{c}c_{m}^{2} \beta]\Psi.
\end{equation}

{\bf 8. Conclusion} \vskip 8pt

In conclusion, we think there may be two kinds of motion phenomena
in nature: One is lowerlumial phenomena $(0\leq v<c)$, the other
is superlumial phenomena $(c\leq v<c_{m})$. In this paper, we
research the classical and quantum theory for superlumial
particle, which ia based on two postulates: The principle of
relativity and the universal of maximum velocity. We obtain all
the results for superluminal theory should be tested by the future
experiment.

\newpage
\end{document}